\documentclass[superscriptaddress,nofootinbib,twocolumn,amsmath,amssymb,aps,prl]{revtex4-2}

\usepackage{xcolor}
\usepackage{braket}
\usepackage{graphicx}
\usepackage{dcolumn}
\usepackage{bm}
\usepackage[mathlines]{lineno}
\usepackage[english]{babel}
\usepackage{enumerate}
\usepackage{comment}
\usepackage{enumitem}

\newcommand{\red}[1]{{\color{red}}}

\begin{document}

\def\be{\begin{equation}}
\def\ee{\end{equation}}
\def\bea{\begin{eqnarray}}
\def\eea{\end{eqnarray}}

% Definizione dei comandi
\newtheorem{theo}{Theorem}
\newtheorem{theor}{Theorem}
\newtheorem{cor}{Corollary}
\newtheorem{lem}{Lemma}
\newtheorem{prop}{Proposition}
\newtheorem{ins}{Insight}
\newtheorem{alg}{Algorithm}

\newtheorem{defin}{Definition}
\newtheorem{ass}{Assumption}
\newtheorem{rem}{Remark}

\newcommand{\qed}{$\blacksquare$}

% -------------------------------

\title{Topology identification of autonomous quantum dynamical networks}

\author{Stefano Gherardini}
\thanks{These authors contributed equally to this work}
\address{CNR-INO, Area Science Park, Basovizza, I-34149 Trieste, Italy}
\address{Department of Physics and Astronomy \& LENS, University of Florence, 50019 Sesto Fiorentino, Italy.}
\affiliation{SISSA, via Bonomea 265, 34136 Trieste, Italy}
\email{stefano.gherardini@ino.cnr.it}

\author{Henk J. van Waarde}
\thanks{These authors contributed equally to this work}
\address{Bernoulli Institute for Mathematics, Computer science and Artificial intelligence, University of Groningen, The Netherlands.} 
%\address{Automatic Control Laboratory, ETH Z\"{u}rich, Physikstrasse 3, 8092 Z\"{u}rich, Switzerland.}

\author{Pietro Tesi}
\address{Department of Information Engineering, University of Florence, via di Santa Marta 3, 50139 Florence, Italy.}

\author{Filippo Caruso}
\address{Department of Physics and Astronomy \& LENS, University of Florence, 50019 Sesto Fiorentino, Italy.}

\begin{abstract}
Topology identification comprises reconstructing the interaction Hamiltonian of a quantum network by properly processing measurements of its density operator within a fixed time interval. It finds application in several quantum technology contexts, ranging from quantum communication to quantum computing or sensing. In this paper, we provide analytical conditions for the solvability of the topology identification problem for autonomous quantum dynamical networks. The solvability condition is then converted in an algorithm for quantum network reconstruction that is easily implementable on standard computer facilities. The obtained algorithm is tested for Hamiltonian reconstruction on numerical examples based on the quantum walks formalism.      
\end{abstract}

\date{\today}

\maketitle

\section{I. Introduction}

A quantum dynamical network is a complex structure, with non-trivial connectivity, composed by spatially separated quantum mechanical systems or nodes with quantum features that are locally stored and manipulated\,\cite{book_quantum_info2000}. 

Studies on quantum networks have several motivations: First of all, there is a growing interest in realizing quantum communication backbones, able to connect remote nodes that are linked by quantum channels\,\cite{MattlePRL1996,CiracPRL1997,DuanNature2001,AspelmeyerIEEE2003,PirandolaNature2016,WehnerScience2018,CacciapuotiIEEE2019,EckerPRX2019}. Secondly, it is worth recalling both the important role of quantum effects in energy transport problems that are popular in many fields of science, as physics, biology, chemistry and information science\,\cite{SchwartzNature2007,LeeScience2007,EngelNature2007,ColliniNature2010,VicianiPRL2015,VicianiSciRep2016,CarusoMaze2016,GehringNRP2019}, as well as relevant applications of quantum networks in many-body physics and quantum computing\,\cite{HarrisNatPhot2017,RoushanScience2017,ChenuPRL2017,KingNaturePRL,AruteNature2019,HudomalCP2020}. 

In all these different contexts, the topology of the networks\,\cite{BookSherTopology2002} is a key aspect that highly influences time-changes of their state. In fact, time behaviors are determined by the interconnection between the nodes of the network, and this is true for both the classical and quantum cases. For example, in the case of consensus dynamical networks, which require the agreement of internal processes on a single data value, the network reaches consensus if and only if the interconnection matrix obeys certain properties\,\cite{Olfati_SaberIEEE2007,MazzarellaTAC2015,ShiTAC2016}. It is thus desirable to determine procedures, both analytical and numerical, that allows for the characterization of the geometric properties of a dynamical network constituted by distinct and independent subsystems. 

Thus, in all cases where network topology is unavailable or uncertain, \emph{topology identification} aims to determine the network structure and possibly the weights of the links between the nodes of the network, by using measurements of its state evolution\,\cite{MaterassiTAC2010}. In doing this, we thus assume that the states of the network, or part of them, can be measured.  

For networks whose nodes exhibit non-quantum behaviors, several topology identification techniques have been already developed, both in the deterministic and stochastic case. In this regard, it is worth mentioning the inverse covariance estimation methods\,\cite{HassanIEEE2016,MorbidiSCL2014}, techniques based on power spectral analysis\,\cite{ShahrampourTAC2015}, compressive sensing\,\cite{SanandajiACC2011,MaterassiSCL2013}, but also topology reconstruction via a transfer matrix of the network\,\cite{GoncalvesTAC2008,WuTCNS2016} or network state matrices using constrained Lyapunov equations\,\cite{vanWaardeTAC2019}.

In this paper, we specifically address the problem of identifying the topology of autonomous quantum dynamical networks. The latter are provided by $d$-dimensional quantum closed systems that, once initialized, evolve according to a time-independent Hamiltonian. In such a case, one is allowed to talk in general about \emph{Hamiltonian reconstruction}. Instead, if a network structure can be identified, the Hamiltonian is provided by the sum of both the Hamiltonian of each quantum network subsystem, and interaction terms taking into account the coupling between them. Here, the aim of the topology identification is to understand the presence of interaction couplings and their intensity, corresponding one-to-one to the topology of the analysed network.

In this scenario, the following three main results are discussed, which may be regarded as a generalization to quantum systems of the results in\,\cite{vanWaardeTAC2019}:\\
(i) We provide sufficient conditions that ensure the solvability of topology identification problems for autonomous quantum networks, by measuring the complete state of the network in the time instants within the interval $[0,\tau]$ with $\tau>0$. \\
(ii) We propose an algorithm to infer network topology for generic autonomous quantum network dynamics undergoing unitary evolution. The algorithm is based on the obtained solvability condition and is designed to be implementable on standard computer facilities. \\
(iii) Sufficient conditions for the solvability of the identification problem are given also in the case of measurements with \emph{partial information}, namely not containing all the values of the network state at each time instant. These results rely on fulfilling observability conditions through the partial measurement of the network states in different time instants. 

The paper is organized as follows. 
%In Sec.\,\ref{sec:DQN}
In Sec.\,II we introduce the basic mathematical elements that define a dynamical quantum network. In addition, we provide the tools to interpret an arbitrary many-body quantum system as a quantum network. 
%Then, in Sec.\,\ref{sec:prob_formulation}, 
Then, in Sec.\,III, we formulate the topology reconstruction problem by using measurements of the network density operator within the time interval $[0,\tau]$. Subsequently, 
%in Sec.\,\ref{sec:top_id_full_info}, 
in Sec.\,IV, we analyse under which analytical conditions the considered topology identification problem, applied to autonomous quantum networks, is solvable by using measurements of the full network density operator within $[0,\tau]$. Specifically, we prove a theorem that provides a sufficient condition for the solvability of the identification problem by simply resolving an algebraic commutation relation. Once determined the conditions for the solvability of the identification problem, 
%in Sec.\,\ref{sec:q_network_rec} 
in Sec.\,V we provide an algorithm, which can be %easily 
implemented on standard computer facilities. The analytical results are tested numerically for Hamiltonian reconstruction on a model based on the quantum random walk formalism. Moreover, in Sec.\,VI, we give preliminary results on topology identification when the network state is only partially measured. Specifically, we show that it is possible to reconstruct the Hamiltonian of an autonomous quantum network if the latter is observable (in a control-theoretic sense) and we can measure the diagonal elements of its density operator by initializing the network on linearly independent initial states in different runs. Finally, a discussion of the results and some outlooks (Sec.\,VII) concludes the paper.

\section{II. Dynamical quantum networks}\label{sec:DQN}

Let us introduce a dynamical quantum network as a quantum system with total dimension $d$ that is the collection of quantum subsystems. According to the laws of quantum mechanics, if the quantum system is an isolated system, its dynamics is assumed \emph{unitary}. Thus, its state vector $|\psi_t\rangle \equiv \boldsymbol{\psi}_t = (\psi_{1},\ldots,\psi_{d})^{T} \in \mathbb{C}^{d}$ is propagated over time by the linear \emph{unitary} operator $\mathcal{U}$ such that
\begin{equation}\label{eq:psi_t}
\boldsymbol{\psi}_t = \mathcal{U}_{t,0}\,\boldsymbol{\psi}_0
\end{equation}
with $\boldsymbol{\psi}_0 = \boldsymbol{\psi}_{t_0}$. Along with the state vector $\boldsymbol{\psi}_t$, we can also introduce the density operator (denoted as $\rho_t$) of the network. It is defined by the outer product of $\boldsymbol{\psi}_t$, i.e., $\rho_t \equiv \boldsymbol{\psi}_{t}\boldsymbol{\psi}_{t}^{\dagger}$ where $(\cdot)^{\dagger}$ stands for the conjugate transpose (or Hermitian transpose) of the operator $(\cdot)$. The density operator provides the statistical description of the state of any quantum system, whereby, in the specific case of a quantum network, return both the quantum description of each network subsystem and all the interference patterns (thus, non-zero quantum correlations) between them. In this regard, it is worth noting that the latter originates only if the nodes of the network are interacting subsystems. As known from quantum mechanics principles\,\cite{Sakurai1994,BreuerBook}, at any time instant $t$ the density operator $\rho_t$ must obey the following constraints: $\rho_t$ is (i) an Hermitian operator, namely $\rho_t^{\dagger}=\rho_t$\,; (ii) positive semi-definite, i.e., given a generic quantum state $\boldsymbol{\phi}$, $\boldsymbol{\phi}^{\dagger}\rho_{t}\,\boldsymbol{\phi} = |\boldsymbol{\phi}^{\dagger}\boldsymbol{\psi}_t|^2 \geq 0$\,; (iii) is trace-preserving, i.e., ${\rm Tr}[\rho_t]=1$. Among the consequences, this implies that the elements along the diagonal of $\rho_t$ are always real numbers summing to $1$. Moreover, the time-evolution of the density operator follows the so-called Liouville–von Neumann equation for any $t\in[0,\infty)$, i.e.,
\begin{equation}\label{eq:LvNE}
    \dot{\rho}_t = \frac{d}{dt}\rho_t = -\frac{i}{\hbar}[H,\rho_t]
\end{equation}
that returns the solution $\rho_t = \mathcal{U}_{t,0}\rho_{0}\,\mathcal{U}_{t,0}^{\dagger}$. In Eq.\,(\ref{eq:LvNE}), $H$ is the Hamiltonian of the network, $[K,J]$ denotes the commutator between the operators $K$ and $J$: $[K,J]=KJ-JK$, and $\hbar$ is the reduced Planck constant. By means of Eq.\,(\ref{eq:LvNE}), one is able to propagate (over time) the initial state $\rho_0$ ensuring that all its properties (i)-(iii) are automatically fulfilled. By solving Eq.\,(\ref{eq:LvNE}) as a function of $\boldsymbol{\psi}_t$, the Schr\"{o}dinger equation $i\hbar\,\dot{\boldsymbol{\psi}}_t = H\boldsymbol{\psi}_t$ is recovered, as well as the unitary dynamics (\ref{eq:psi_t}).

\begin{rem}\label{remark_linear_equation}
Instead of solving the Liouville-von Neumann equation (\ref{eq:LvNE}) to get the quantum network dynamics, it may useful to work with the column vector $\boldsymbol{\lambda}_t$, which is obtained by vectorizing the density operator $\rho_t$: 
\begin{eqnarray}
&\boldsymbol{\lambda}_t \equiv {\rm vec}[\rho_t]&\nonumber \\
&= (\rho_t^{(11)},\ldots,\rho_t^{(d1)},\rho_t^{(12)},\ldots,\rho_t^{(d2)},\ldots,\rho_t^{(dd)})^T \in\mathbb{C}^{d^2 \times 1}&.\nonumber \\
&&
\end{eqnarray}
In this way, according to the super-operator formalism \cite{HavelJMP2003}, the Liouville-von Neumann equation (\ref{eq:LvNE}) can be written as a linear differential equation in the column vector $\boldsymbol{\lambda}_t$:
\begin{equation}\label{eq:linearEq}
\dot{\boldsymbol{\lambda}}_t = \mathcal{L}\,\boldsymbol{\lambda}_t\,\,\,\,\,
\text{namely}\,\,\,\,\,\boldsymbol{\lambda}_t = e^{\mathcal{L}(t-t_{0})}\boldsymbol{\lambda}_0
\end{equation}
where $\mathcal{L} = -\frac{i}{\hbar}(\mathbb{I}_{d} \otimes H - H^T \otimes \mathbb{I}_d) \equiv -\frac{i}{\hbar}\widetilde{H}\in\mathbb{C}^{d^2\times d^2}$, $\mathbb{I}_{d}$ is the identity matrix of size $d$ and $\otimes$ denotes the Kronecker product. By construction, $\mathcal{L}$ is a skew-Hermitian (or anti-Hermitian) operator meaning that $\mathcal{L}^{\dagger}$+$\mathcal{L}=0$. \\
The linear formulation of quantum network dynamics will be adopted in Sec.\,VI, where we will address the solvability of topology identification problems with partial information.
\end{rem}

\subsubsection{Quantum networks from many-body systems}

For practical applications in quantum communication and computing, a quantum network is usually constituted by $N$ interacting $\ell$-levels quantum systems, each of them characterized by its own dynamics and corresponding to a node of the quantum network. Thus, in this subsection we briefly discuss how one can deal with the coupling structure of a many-body system composed of interacting subsystems and, then, express the time-evolution of the corresponding density operator via an instance of the Liouville-von Neumann equation (\ref{eq:LvNE}). In this way, the identification of the unknown interaction structure of a quantum many-body system can be achieved by addressing the reconstruction of the Hamiltonian in (\ref{eq:LvNE}), as we have formulated below.

Under the hypothesis that a quantum network is provided by $N$ $\ell$-levels quantum systems, the dimension of the network is equal to $d=\ell^{N}$, and, concerning the network Hamiltonian, the contributions associated to each node of the network can be decoupled by the ones of the coupling terms. Formally,
\begin{equation}\label{eq:H_total}
    H = H_0 + H_{\rm int} = \sum_{k=1}^{N}\omega_{k}H_{k} + H_{\rm int}
\end{equation}
where $k$ is the index over the network nodes. It is worth noting that a different characteristic frequency $\omega_{k}$ and a local Hamiltonian $H_k$ is associated to each node, while $H_{\rm int}$ denotes the interaction Hamiltonian. All the operators $H_k$, for $k=1,\ldots,N$, and $H_{\rm int}$ are Hermitian with size $d$.

In the quantum many-body systems scenario, it is common practice to assume two-body interactions and compositions of them at any instant of time $t$. This means that the interaction Hamiltonian $H_{\rm int}$ can be further decomposed in a sum of $D \equiv N^2 - N = N(N-1)$ coupling terms $A_{k}A_{j}$, with $k \neq j$, each of them corresponding to one specific link. This means that
\begin{equation}\label{eq:H_int_two-body_ints}
    H_{\rm int} = \sum_{k,j=1;\,k \neq j}^{N} \alpha_{k,j}A_{k}A_{j}\,.
\end{equation}
Hence, by substituting (\ref{eq:H_int_two-body_ints}) in (\ref{eq:H_total}), it holds that the Liouville-von Neumann equation of a quantum many-body system (written as a dynamical network) is given by
\begin{equation}\label{eq:LvNE_many_body}
  \dot{\rho}_t = -\frac{i}{\hbar}\left[(H_0 + \boldsymbol{\alpha}^{T}C\boldsymbol{\alpha})\right] = R_0(t) -\frac{i}{\hbar}\left[\boldsymbol{\alpha}^{T}C\boldsymbol{\alpha},\rho_t\right]
\end{equation}
where $R_0(t) \equiv -i[H_0,\rho_t]/\hbar$ is known and completely determined by the knowledge of the local dynamics of the network, while $\boldsymbol{\alpha}$ and $C$ are provided by the relations
\begin{eqnarray*}
    &&\boldsymbol{\alpha} = \left(\sqrt{\alpha_{1,2}}\,\mathbb{I}_{d}\,,\ldots,\sqrt{\alpha_{N,N-1}}\,\mathbb{I}_{d}\right)^{T}\in\mathbb{C}^{Dd\times d}\\
    && C = {\rm diag}\left(\left\{A_{k}A_{j}\right\}\right)\in\mathbb{C}^{D(d\times d)}.
\end{eqnarray*}
Further details can be found in the Appendix A, where we also provide a microscopic derivation of a quantum network by using the many-body formalism and writing each operator $A_k$, $k=1,\ldots,N$, as a function of a complete orthornormal basis of eigenoperators. 

\begin{rem}
The complete structure of the operator $C$ (sparse Hermitian matrix) is defined by the laws of quantum mechanics ruling two-body interactions. This means that the only unknown quantity to be reconstructed, returning the topology of the network, is the vector $\boldsymbol{\alpha}$ that contains all the interaction couplings.
\end{rem}

\section{III. Problem formulation}\label{sec:prob_formulation}

Let us consider a dynamical quantum network with unitary dynamics described by the Liouville-von Neumann equation (\ref{eq:LvNE}), and assume that the system Hamiltonian $H$ is not directly available. During the time-interval $[0,\tau]$, indeed, one can have access only to the density operator $\rho_t$ or a portion of it.

In the following, we will address the problem of identifying the topology of a generic autonomous quantum dynamical network, by following the prescription of \emph{network reconstruction} problems as given in the engineering literature. Specifically, the network reconstruction problem involves the identification of the exact value of the Hamiltonian $H$ of the quantum network (it corresponds to infer the vector $\boldsymbol{\alpha}$ of interaction couplings if one has knowledge of the quantum many-body structure) on the basis of measurements of (part of) the density operator $\rho_t$ or equivalently of the ensemble vector $\boldsymbol{\lambda}_t$. 

We will provide analytical conditions under which this important prerequisite holds, and we will present an algorithm for the Hamiltonian reconstruction problem.

\section{IV. Topology identification with full information}\label{sec:top_id_full_info}

In this section, we provide analytical conditions that ensure the solvability of the topology identification problem for autonomous quantum networks by measuring the full density operator $\rho_t$ within the time-interval $[0,\tau]$.

For this purpose, let us consider the density operator $\rho_t$ for any $t\in[0,\tau]$ and take the network Hamiltonian $H$ as a $d$-dimensional Hermitian operator $\in\mathbb{C}^{d \times d}$. Moreover, we also assume that $H$ has zero diagonal entries that helps focusing our derivations for the identification of the topology (i.e., of the links among the nodes) of an autonomous quantum network. Any matrix $M$ with these two properties (i.e., Hermitianity and zero diagonal entries) is here denoted as \emph{admissible}, and the set of all admissible matrices as 
\begin{equation}
\mathcal{A} \equiv \{ M \in \mathbb{C}^{d \times d} \mid M = M^\dagger \text{ and } M_{ii} = 0, \:\: \forall i = 1,\ldots,d \}.
\end{equation}
By running experiments based on Eq.\,\eqref{eq:LvNE}, we get a specific trajectory $\rho_t$, with $t \in [0,\tau]$, for any given initial state $\rho_0$. Thus, the set of admissible matrices that allow for such data is given by
\begin{equation}
\mathcal{A}_\rho \equiv \left\{M \in \mathcal{A} \mid \dot{\rho_t} = -\frac{i}{\hbar} [M,\rho_t] \:\: \forall t \in [0,\tau] \right\}.
\end{equation}
In addition, let us define the solvability condition of the topology identification problem: 
\begin{defin}
Let $\rho_t$ be given for $t \in [0,\tau]$. Then, the topology identification problem is called \emph{solvable} if $\mathcal{A}_\rho = \{H\}$.
\end{defin}

Now, by integrating Eq.\,(\ref{eq:LvNE}) and taking the network Hamiltonian as an admissible operator, we provide a \textit{sufficient condition} for the solvability of the aforementioned identification problem in the form of the following theorem.

\begin{theo}\label{theorem_1}
Define the matrices
\begin{eqnarray}
&P \equiv \int_{0}^{\tau}\rho_{t}\,dt&\label{eq:def_P} \\
&Q \equiv i\hbar\,(\rho_{\tau} - \rho_{0}).&\label{eq:def_Q}
\end{eqnarray}
with $\tau>0$ arbitrary. Then, the topology identification problem is solvable if there exists a {\bf unique} admissible $\hat{M} \in \mathcal{A}$ that satisfies the relation
\begin{equation}\label{skew-Lyapunov}
[\hat{M},P] = Q \,.
\end{equation}
\end{theo}

The proof of Theorem \ref{theorem_1} is in Appendix B. As it will be shown later, for a given choice of $\rho_{0}$ and $\tau$, the knowledge of the density operator $\rho_t$ in a set of time instants $t\in[0,\tau]$, whose number increases at least linearly with the number $d$ of quantum network's nodes, may be sufficient to solve the topology identification problem. However, the larger is the number of time instants at which $\rho_t$ is evaluated (for a fixed value of $\tau$) the smaller is the topology reconstruction error.

\begin{rem}
For the purpose of topology identification, let us observe once more that taking a quantum many-body system as the quantum network is an instance of the more general problem that we have previously formulated. In fact, by integrating Eq.\,(\ref{eq:LvNE_many_body}) within $[0,\tau]$ and defining the matrices $P \equiv \int_{0}^{\tau}\rho_{t}\,dt$ and 
\begin{equation*}
Q \equiv i\hbar\,(\rho_{\tau} - \rho_{0}) - [H_0,P]\,,
\end{equation*}
the problem of identifying the interaction Hamiltonian $H_{\rm int}=\boldsymbol{\alpha}^{T}C\boldsymbol{\alpha}$ is again solvable if there exists a {\bf unique} admissible $\hat{M} \in \mathcal{A}$ that satisfies $[\hat{M},P] = Q$.  
\end{rem}

In Theorem \ref{theorem_1} the matrices $P$ and $Q$ can be computed from data. Thus, in terms of topology identification, we conclude that solving Eq.\,\eqref{skew-Lyapunov} for $\hat{M} \in \mathcal{A}$ leads to an effective method to reconstruct $H$ from measurements of $\rho_t$ for $t\in[0,\tau]$. However, one also needs to address the issue of determining under which conditions there exists a {\bf unique} solution $\hat{M}\in\mathcal{A}$ to Eq.\,\eqref{skew-Lyapunov}. In this regard, a \emph{necessary and sufficient condition} for the uniqueness of $\hat{M}\in\mathcal{A}$ is provided by the following proposition, whose proof is in Appendix C.

\begin{prop}\label{proposition_1}
There exists a unique $\hat{M}\in\mathcal{A}$ satisfying \eqref{skew-Lyapunov} \emph{if and only if} the zero matrix (operator with all terms equal to zero and denoted as $\emptyset$) is the only element of $\mathcal{A}$ that commutes with $P$.
\end{prop}

To Proposition \ref{proposition_1}, the following corollary can be associated. 

\begin{cor}\label{corollary_1}
A necessary condition for the uniqueness of the nonzero solution $\hat{M}\in\mathcal{A}$, resulting by solving the equation $MP - PM = Q$, is that $\hat{M}$ does not commute with $P$.
\end{cor}

Corollary\,\ref{corollary_1} is a direct consequence of Proposition \ref{proposition_1}, and can be used as a preliminary check to evaluate whether the solution $\hat{M}\neq 0$ of the topology identification problem may be effectively unique. In fact, if $[\hat{M},P]=0$, then $\hat{M}$ is definitely not unique. Moreover, if $Q \neq \emptyset$, then from Proposition \ref{proposition_1} and Corollary\,\ref{corollary_1}, it is guaranteed the existence of a unique admissible solution $\hat{M} \neq \emptyset$ that solves the equation $MP - PM = Q$ with $P$ and $Q$ computed from data. From these theoretical results, we can put in place an effective strategy to solve the topology identification problem here considered. This strategy relies in measuring the density operator $\rho_t$ of the network within the time interval $[0,\tau]$ until the equation $MP - PM = Q$ has nonzero solution $\hat{M} \in \mathcal{A}$ that does not commute with $P$. A reconstruction algorithm will be thus presented in the next section.

\section{V. An algorithm for quantum network reconstruction}\label{sec:q_network_rec}

In this section, we provide an algorithm for the reconstruction of the quantum network Hamiltonian that requires, as input data, measurements of all the elements of $\rho_t$. In this regard, let us recall that: (i) the reconstruction problem is solvable \emph{if} there exists a unique matrix $\hat{M}\in\mathcal{A}$ that satisfies the relation $[\hat{M},P]=Q$ with $P$, $Q$ computed from measured data (Theorem \ref{theorem_1}); (ii) there exists a unique $\hat{M}\in\mathcal{A}$ obeying $[\hat{M},P]=Q$ \emph{if and only if} the zero matrix $\emptyset$ is the only element of $\mathcal{A}$ that commutes with $P$ (Proposition \ref{proposition_1}). Conditions (i) and (ii) are the guidelines to formulate the reconstruction algorithm. Theorem\,\ref{theorem_1} and Proposition\,\ref{proposition_1}, indeed, imply that if the reconstruction of the network Hamiltonian (here we are interested only to the interaction components, i.e., to the network topology) is solvable, then the solution to this problem can be obtained by determining the unique Hermitian operator with zero diagonal elements that obeys to Eq.\,\eqref{skew-Lyapunov}. 

One way to resolve the matrix equation $[M,P]=Q$, under the constraint of $M$ Hermitian operator, is to vectorize the matrix equation by writing a standard system of \emph{linear} equations, and then impose the Hermitian symmetry of $M$ by means of \emph{linear} constraints. Specifically, by defining the operator
\begin{equation}
\widetilde{P} \equiv \left( P^{T} \otimes \mathbb{I}_{d} - \mathbb{I}_{d} \otimes P \right) \in \mathbb{C}^{d^2 \times d^2}
\end{equation}
and reshaping the matrices $M$ and $Q$ in column vectors of dimension $d^2$, denoted respectively as $\boldsymbol{m} \equiv {\rm vec}[M]$ and $\boldsymbol{q} \equiv {\rm vec}[Q]$, the matrix equation $[M,P]=Q$ can be recast in the linear equation
\begin{equation}\label{eq:linear_equation_alg_1}
\widetilde{P}\boldsymbol{m}=\boldsymbol{q}.
\end{equation}
Then, instead of resolving the equation $\widetilde{P}\boldsymbol{m}=\boldsymbol{q}$ as a function of $\boldsymbol{m}$ with the constraint that the resulting solution satisfies the property of Hermitian symmetry, we write an \emph{enlarged} linear system such that the required constraints (Hermitian symmetry and solution with diagonal elements equal to zero) are automatically fulfilled. For this purpose, we define the matrices $F_1 \in \mathbb{R}^{d\times d^2}$ and $F_2 \in \mathbb{C}^{\frac{d(d-1)}{2} \times d^2}$. The former has almost all terms equal to zero except to the $(k+1,kd+1)$-th elements, with $k=0,\ldots,d-1$, that are all equal to $1$. This guarantees that, by imposing $F_{1}\boldsymbol{m}=\emptyset$, all the diagonal elements of $M$ (corresponding to the $(kd+1,kd+1)$-th diagonal elements of $M$ with $k=0,\ldots,d-1$) are equal to zero. Instead, $F_2$ is designed to ensure that $M=M^{\dagger}$. For the sake of a clearer presentation, here we provide a simplified expression of $F_2$ under the hypothesis that $M$ is a matrix with real elements. In such a case, one can easily check that the validity of the relation $F_{2}\boldsymbol{m}=\emptyset$ is equivalent to ensure that $M=M^{T}$, where
\begin{equation*}
F_{2,kj}\equiv\begin{cases}
+1,\,\,\,\,\text{if}\,\,\,\,j=(\ell-1)d + i \\
-1,\,\,\,\,\text{if}\,\,\,\,j=(i-1)d + \ell \\
0,\,\,\,\,\text{otherwise}
\end{cases}
\end{equation*}
with $k=1,\ldots,d(d-1)/2$, $\ell=1,\ldots,d-1$ and $i=\ell +1,\ldots,d$. A similar, though more involved, expression holds in case $M$ is a matrix of complex numbers. In conclusion, the enlarged linear system that recasts the matrix equation $[M,P]=Q$, with $M$ Hermitian and diagonal elements equal to zero, is just given by
\begin{equation}
\widetilde{P}'\boldsymbol{m} = \boldsymbol{q}'\,\,\,\,\text{where}\,\,\,\,\widetilde{P}'\equiv\begin{bmatrix}
\widetilde{P} \\ F_{1} \\ F_{2}
\end{bmatrix}\,\,\,\,\text{and}\,\,\,\,\boldsymbol{q}'\equiv\begin{bmatrix}
\boldsymbol{q} \\ \emptyset \\ \emptyset
\end{bmatrix}.
\end{equation}
Accordingly, the algorithm for the reconstruction of the quantum network Hamiltonian, using measurements of the whole density operator $\rho_t$, is defined as follows.
\begin{alg}\label{alg:alg1}
Network Hamiltonian reconstruction with full information:
\begin{eqnarray*}
&&\text{\rm If}\,\,\,\,{\rm rank}(\widetilde{P}')= d^2
\,\,\,\Longrightarrow\,\,\,
\hat{\boldsymbol{m}}=(\widetilde{P}')^{+}\boldsymbol{q}'\,; \\
&&\text{\rm otherwise, $[M,P] = Q$ has non-unique solutions $\hat{M}\in\mathcal{A}_\rho$}.
\end{eqnarray*}
$(\widetilde{P}')^{+} \equiv \widetilde{P}'^{*}(\widetilde{P}'\widetilde{P}'^{*})^{-1}$ denotes the Moore-Penrose right inverse of $\widetilde{P}'$ with $\widetilde{P}'^{*}$ the corresponding conjugate transpose. Then, $\hat{M}$ is obtained by applying the inverse of the vectorization operation, thus reshaping the column vector $\hat{\boldsymbol{m}}$ in a square Hermitian operator with diagonal elements equal to zero.
\end{alg}
It is worth observing that asking for ${\rm rank}(\widetilde{P}')= d^2$ is the necessary requirement for the uniqueness of the solution $\hat{M}\in\mathcal{A}_\rho$ solving the matrix equation $[M,P] = Q$. However, this does not imply that a value of $\hat{\boldsymbol{m}}=(\widetilde{P}')^{+}\boldsymbol{q}'$ cannot be obtained from the calculation, albeit in this stage we do not have control on the magnitude of the reconstruction error. Moreover, Algorithm \ref{alg:alg1} has the advantage of determining its solution by solving a system of linear equations (i.e., $\widetilde{P}'\boldsymbol{m} = \boldsymbol{q}'$) that can be achieved using efficient computational tools.

\subsection{Case-study: Quantum random walk model}\label{case_studies}

As a case-study of network Hamiltonian reconstruction, let us consider the quantum random walk model \cite{KempeCP2003,VenegasQIP2012} that is the transposition of the concept of classical random walk to the quantum context. Specifically, we take into account a single walker moving on a graph $G$. The latter is described by the pair $G = (\mathcal{N},\mathcal{E})$, where $\mathcal{N}$ denotes the set of nodes (or vertices) of the graph and $\mathcal{E}$ is the set of links that couple pairs of nodes $(\mathcal{N}_k, \mathcal{N}_{\ell})$. Each node is associated to a different walker position, while the links correspond to the probability that the walker jumps from a node to another. The links belonging to $\mathcal{E}$ can be summarized in the adjacency matrix $A$, whose elements are given by
\begin{equation*}\label{eq:adjecency_matrix}
    A_{kj} =
    \begin{cases}
    1,\,\,\,\,\text{if}\,\,\,\,(\mathcal{N}_k, \mathcal{N}_j)\in \mathcal{E}  \\
    0,\,\,\,\,\text{if}\,\,\,\,(\mathcal{N}_k, \mathcal{N}_j)\neq \mathcal{E}. \\
    \end{cases}
\end{equation*}
Note that here we are implicitly assuming that the links are equally weighted with weights all equal to 1. Moreover, if all the positions of the walker are states with the same energy, then one is allowed to set such energy to a reference constant value. Usually the reference energy value is taken equal to zero, with the result that the Hamiltonian $H$ of the quantum walker is identically equal to the adjacency matrix $A$ ($H = A$).

Also the state of the quantum walker, moving on the graph $G$ with $d$ nodes, is provided by a density operator $\rho_t\in\mathbb{C}^{d\times d}$. The diagonal elements of $\rho_t$ define the probabilities that the walker is in each of the allowed positions (such terms are denoted as \emph{populations}), while the off-diagonal elements are the so-called \emph{quantum coherence} terms that identify interference patterns between nodes. The initial state $\rho_0$ is taken with all the coherence terms equal to $0$ and only one population equal to $1$ (randomly chosen in each realization of the network dynamics).

By addressing the topology identification problem with full information, we measure the elements of the density operator $\rho_t$ within the time interval $t\in[0,\tau]$, and the reconstruction problem consists in determining the exact value of the elements of the adjacency matrix $A$, i.e., in identifying the presence of a link between the nodes of the graph. To validate the performance of the Algorithm \ref{alg:alg1}, we numerically solve the dynamics of $100$ random networks $A$, each of them with an increasing number $d\in[2,30]$ of nodes. The network dynamics is computed with resolution (sampling period) $\delta t = 10^{-2}$. This entails that the total number of samples that compose the quantum network evolution is equal to the ratio $n_s = \tau/\delta t = \tau/10^{-2}$. The random networks are sampled by an Erd\H{o}s-R\'enyi distribution, whereby the nodes of the network are randomly connected and each link is included in the graph with probability $p_{\rm link}$ \cite{Erdos1959,NewmanPRE2001}. We recall that, according to this model, a network with $d$ nodes and $m$ links is sampled with probability
\begin{equation*}
    p_{\rm graph} = p_{\rm link}^{m}(1-p_{\rm link})^{{{d}\choose{2}}-m} 
\end{equation*}
where ${{n_1}\choose{n_2}} \equiv \frac{n_{1}!}{n_{2}!(n_1-n_2)!}$ is the binomial coefficient and $2^{{d}\choose{2}}$ the total number of networks with $d$ nodes. Instead, regarding the initialisation, each time the initial density operator of the network is randomly chosen among one of the states $\rho_0 = \boldsymbol{e}_{k}\boldsymbol{e}_{k}^{T}$ with $k=1,\ldots,d$ (thus, with just one node excited). This choice is the simplest to be realized experimentally. The evolution of the network is then evaluated at discrete time instants within the time interval $[0,\tau]$, where $\tau$ is equal to $1$, $2$ or $3$, all expressed in natural units such that $\hbar$ can be set to $1$. In our simulations, the operator $P$ (it is provided by the integral of $\rho_t$, solution of the system dynamics) within the interval $[0,\tau]$ is approximated as
\begin{eqnarray}\label{eq:approx_P}
P &\equiv& \int_{0}^{\tau}\rho_{t}\,dt \approx \sum_{k=1}^{\widetilde{n}_s}\int_{t_{k-1}}^{t_k}\rho_{t}\,dt\nonumber \\
&\approx& \frac{1}{2}\sum_{k=1}^{\widetilde{n}_s}(t_{k}-t_{k-1})(\rho_{t_{k-1}}+\rho_{t_k}).
\end{eqnarray}
Numerically, the integral $\int_{0}^{\tau}\rho_{t}\,dt$ is computed by assigning different values to $\widetilde{n}_s$, i.e., $n_s/20$, $n_s/10$, $n_s/5$ and $n_s$, with the aim to evaluate how the reconstruction performance of Algorithm \ref{alg:alg1} depends on the resolution used to monitor the network dynamical evolution. Note that smaller is the value of $\widetilde{n}_s$, and fewer evaluations of $\rho_t$ are required. Moreover, for each realization of the simulated dynamics, we introduce a label $\mathfrak{s}$, denoted as \emph{solvability label}, that takes two values: $1$ if the quantum network reconstruction problem is solvable (i.e., $\hat{\boldsymbol{m}}=(\widetilde{P}')^{+}\boldsymbol{q}'$ for ${\rm rank}(\widetilde{P}')= d^2$) with $P$ approximated as in Eq.\,(\ref{eq:approx_P}), and $0$ otherwise, in agreement with the prescriptions of Algorithm \ref{alg:alg1}.

\begin{figure}[t!]%[!htb]
\centering
\includegraphics[scale=0.64]{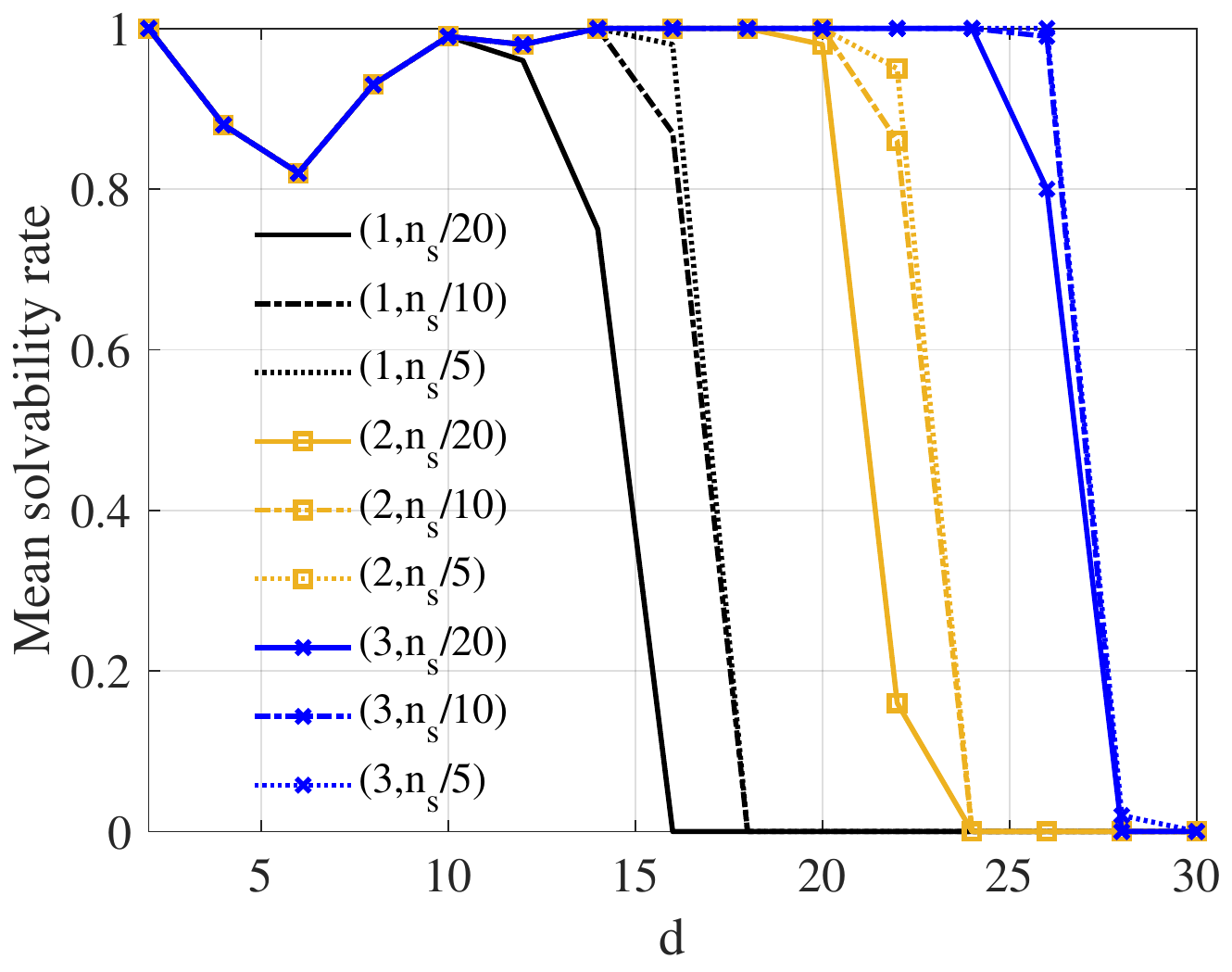}
\caption{Plot of the mean solvability rate $\overline{\mathfrak{s}}$ as a function of $d\in[2,30]$ (number of nodes) and of the pairs $(\tau,\widetilde{n}_s)$, with $\tau\in\{1,2,3\}$ and $\widetilde{n}_s\in\{n_s/20,n_s/10,n_s/5\}$. Specifically, for the considered values of $d$ and $p_{\rm link}=0.5$, the topology identification problem is solved for $100$ random Erd\H{o}s-R\'enyi quantum networks. The black, orange and blue curves refer, respectively, to $\tau=1,2,3$.}
\label{fig:Fig_1}
\end{figure}

In Fig.\,\ref{fig:Fig_1} we plot the mean solvability rate $\overline{\mathfrak{s}}$ in identifying the topology of random Erd\H{o}s-R\'enyi quantum networks as a function of $d\in[2,30]$. The mean solvability rate is equal to the arithmetic mean of $\mathfrak{s}$ obtained in each of the $100$ network reconstructions for every chosen set of parameters, namely
\begin{equation}
\overline{\mathfrak{s}} \equiv \frac{1}{100}\sum_{q=1}^{100}\mathfrak{s}_q\,.
\end{equation}
The value of $\overline{\mathfrak{s}}$ is plotted as a function of both the duration $\tau$ of the network dynamics, and the number of samples $\widetilde{n}_s$ used to calculate numerically the operator $P$. From Fig.\,\ref{fig:Fig_1} the following numerical evidences can be observed: (i) By increasing the number of nodes, linearly larger value of $\tau$ is required to successfully carry out the network Hamiltonian reconstruction problem. This means that, to reconstruct the topology of an Erd\H{o}s-R\'enyi quantum network, we need to monitor its dynamics at least for a linearly longer period of time, independently on the value of $p_{\rm link}$. As an example, for the simulations in Fig.\,\ref{fig:Fig_1} where the value of $\tau$ is fixed and taken equal to $1,2,3$ (corresponding, respectively, to the back, orange and blue curves in the figure), the mean solvability rate $\overline{\mathfrak{s}}=0$ $\forall d>d_{\rm c}$, with $d_{\rm c}$ critical number of nodes depending on $\tau$. For the blue curves, e.g., $d_{\rm c}=28$. However, one can recover $\overline{\mathfrak{s}}=1$ (thus, high-probability topology reconstruction) for $d > d_{\rm c}$ if the value of $\tau$ were increased, linearly with the number of nodes. Moreover, notice that, for quantum networks with few nodes, $\overline{\mathfrak{s}}$ may be smaller than the maximum value $1$ that instead is reached for a larger number of nodes up to the critical value $d_{\rm c}$. For small values of $d$, $\overline{\mathfrak{s}} \neq 1$ since the dynamics of the quantum network in such case are oscillating. Thus, repeated evaluations of the network dynamics at regular discrete times, as in our simulations, provide similar results and thus be not so informative. In conclusion, $\overline{\mathfrak{s}} \neq 1$ can be considered as a \emph{finite-size} effect; in fact, by increasing the dimension of the network, such effect is lost until Algorithm\,\ref{alg:alg1} loses its effectiveness due to the finite and fixed value of $\tau$. (ii) The mean solvability rate $\overline{\mathfrak{s}}$ practically does not depend on the number of samples $\widetilde{n}_s$ used to derive $P$. Hence, the identification problem is solvable also by computing the operator $P$ by means of few measurements of $\rho_t$ within $[0,\tau]$. However, one could expect that, by decreasing the number of measured values of $\rho_t$ (needed for the computation of $P$), the reconstruction error dramatically increases. 
\begin{figure}[t!]%[!htb]
\centering
\includegraphics[scale=0.65]{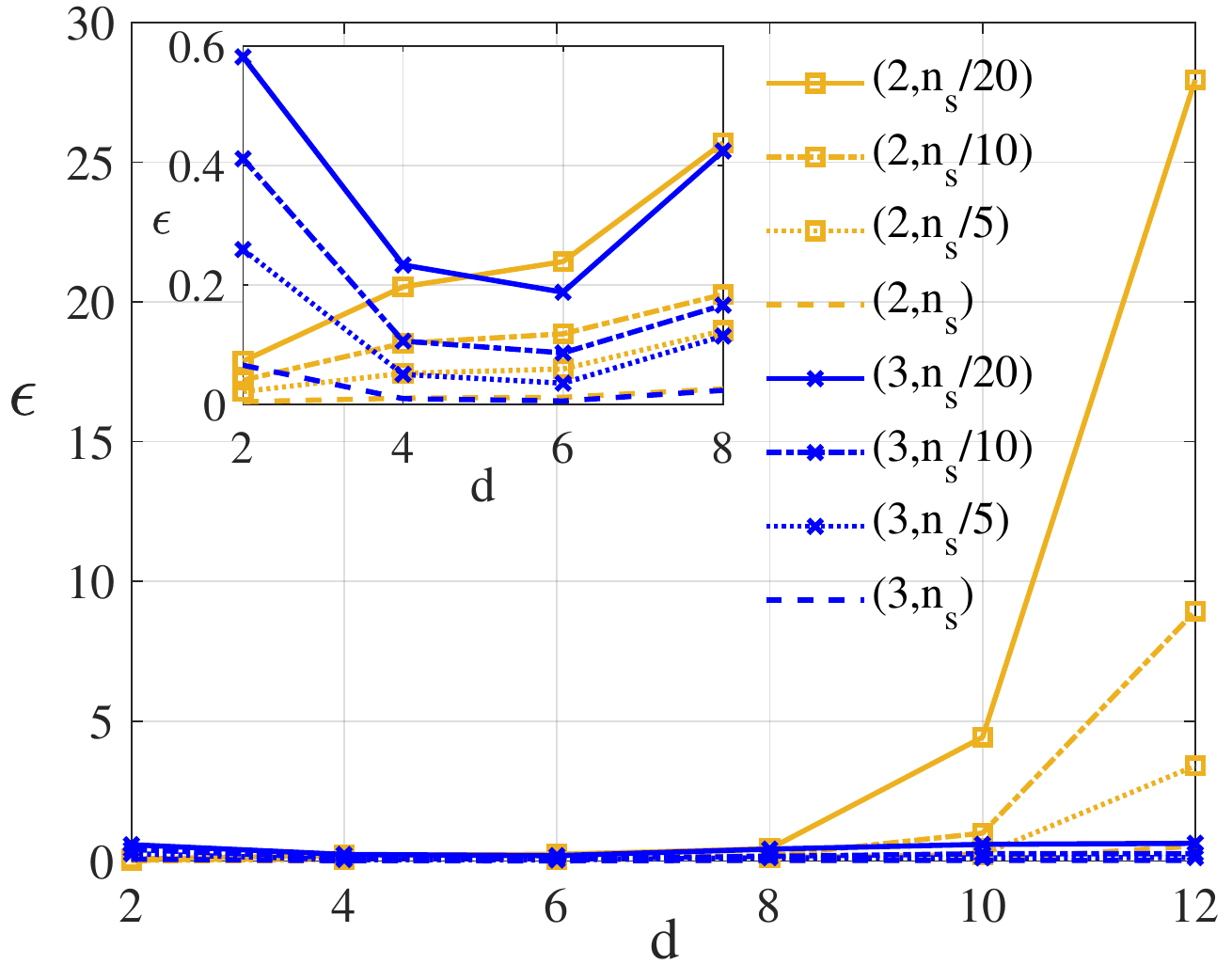}
\caption{Plot of the relative reconstruction error $\epsilon$ in identifying single random Erd\H{o}s-R\'enyi quantum networks with $p_{\rm link}=0.5$. The relative error $\epsilon$ is plotted as a function of $d\in[2,12]$ for different pairs $(\tau,\widetilde{n}_s)$, with $\tau\in\{1,2\}$ and $\widetilde{n}_s\in\{n_s/20,n_s/10,n_s/5,n_s\}$. In the figure, the red and blue curves refer, respectively, to $\tau=2,3$.}
\label{fig:Fig_2}
\end{figure}
This is what has been quantified in Fig.\,\ref{fig:Fig_2}, where, in case the quantum network reconstruction problems are solvable, we plot the relative reconstruction error $\epsilon$ in identifying single random Erd\H{o}s-R\'enyi quantum networks by using full information. The relative reconstruction error is defined as 
\begin{equation}
    \epsilon \equiv \frac{\|\hat{M}-A\|_2}{\|A\|_2}
\end{equation}
where, here, $\left\|\cdot\right\|$ denotes the $L^2$ operator-norm, $A$ is the network adjacency matrix (Hamiltonian) to be identified and $\hat{M}\in\mathcal{A}_{\rho}$ is the reconstructed one returned by the Algorithm \ref{alg:alg1}. In Fig.\,\ref{fig:Fig_2}, the relative error $\epsilon$ is plotted as a function of the number of nodes and for $8$ different pairs $(\tau,\widetilde{n}_s)$, with $\tau\in\{1,2\}$ and $\widetilde{n}_s\in\{n_s/20,n_s/10,n_s/5,n_s\}$. As one can observe (also from the inset), the relative reconstruction error is only slightly greater than $0$ for $d \leq 8$, independently on the value of $\tau$ and $\widetilde{n}_s$. Of course, also in this case, $\epsilon$ is larger if $\widetilde{n}_{s}=n_s/20$, i.e., if only few values of the network density operator are taken into account. Instead, the larger values of $\epsilon$ are observed in Fig.\,\ref{fig:Fig_2} for $d>8$, where the relative reconstruction error remains small for $\tau=3$ and $(\tau=2,\widetilde{n}_s=n_s)$. This means that, despite the identification problem is solvable, the reconstruction error can dramatically increase, especially by decreasing the duration of the network dynamics and using few measurement outcomes to compute $P$.

\section{VI. Solvability of topology identification with partial information}\label{sec:partial_info}

In this section, we give some preliminary results for the situation where the whole network state is not available for measurements. Specifically, we address the solvability of topology identification problems for autonomous quantum dynamical networks by using only measurements of the \emph{diagonal entries} of $\rho_t$, solution of the Liouville–von Neumann equation \eqref{eq:LvNE}. Measuring all the elements of $\rho_t$, indeed, is just a \emph{sufficient condition} to successfully reconstruct the topology of a quantum network in the single realization of its dynamics for a generic initial state $\rho_0$. This means that, by collecting data on $\rho_t$, the reconstruction problem can be solvable also in the case that only a portion of $\rho_t$ is measured.

As stressed in Remark\,\ref{remark_linear_equation}, the Liouville–von Neumann equation can be written as a linear differential equation of the state vector $\boldsymbol{\lambda}_t \in \mathbb{C}^{d^2\times 1}$ by means of the vectorization of $\rho_t$. In this way, one gets $\dot{\boldsymbol{\lambda}}_t = \mathcal{L}\,\boldsymbol{\lambda}_t$ with $\mathcal{L}$ skew-Hermitian operator. For our purposes, let us associate to the dynamical equation $\dot{\boldsymbol{\lambda}}_t = \mathcal{L}\,\boldsymbol{\lambda}_t$ an \emph{output equation} that selects only the diagonal elements of $\rho_t$ at each time $t$. The output equation is 
\begin{equation*}
    \boldsymbol{y}_t = C\boldsymbol{\lambda}_t \in \mathbb{C}^{d\times 1}
\end{equation*}
with $C \in \mathbb{R}^{d\times d^2}$ defined as
\begin{equation*}
C_{kj}=\begin{cases}
1,\,\,\,\,\text{if}\,\,\,\,j=(k-1)d + k \\
0,\,\,\,\,\text{otherwise}
\end{cases}
\end{equation*}
where $d$ is the dimension of the quantum network. It is worth noting that the dynamical and output equations are fully characterized by the initial state $\boldsymbol{\lambda}_{0}$ and the pair $(C,\mathcal{L})$. In what follows we will resort to the concept of \emph{observability} \cite{Kalman1963}. The pair $(C,\mathcal{L})$ is observable if
\begin{equation}
{\rm rank}\begin{bmatrix} C \\ C\mathcal{L} \\ \vdots \\ C\mathcal{L}^{d^2 - 1} \end{bmatrix}=d^2\,.
\end{equation}
For the sake of clarity, the output trajectory $\boldsymbol{y}_t$ in the time interval $[0,\tau]$, resulting from the initial state $\boldsymbol{\lambda}_{0}$, is denoted as $\boldsymbol{y}_{\lambda_0}(t)$.

Under the assumption that the quantum network is observable, we provide (in the following Proposition) a condition under which the operator $\mathcal{L}$ can be \emph{uniquely} identified from \emph{multiple} partial measurements of $\rho_t$ with $C$ fixed a-priori. 

\begin{prop}\label{proposition_2}
Let $\boldsymbol{\lambda}_{0}^{(\ell)}$ be $d^2$ linearly independent initial states for $\ell = 1,\ldots,d^2$. Let us also assume to measure the output equation $\boldsymbol{y}_t$ of the quantum network dynamics that is obtained by initializing the quantum network in each of the $d^2$ input states, thus having access to $\boldsymbol{y}_{\lambda_0^{(\ell)}}(t)$ for $t \in [0,\tau_{\ell}]$, $\tau_{\ell}>0$ and $\ell=1,\ldots,d^2$. If the pair $(C,\mathcal{L})$ is observable, then the operator $\mathcal{L}$ is {\bf uniquely} identifiable from $[\boldsymbol{y}_{\lambda_0}^{(1)},\ldots,\boldsymbol{y}_{\lambda_0}^{(d^2)}]$. In other words, $\overline{\mathcal{L}}=\mathcal{L}$ for any pair $(C,\overline{\mathcal{L}})$ that generate the outputs $[\boldsymbol{y}_{\lambda_0}^{(1)},\ldots,\boldsymbol{y}_{\lambda_0}^{(d^2)}]$.
\end{prop}

The proof of Proposition\,\ref{proposition_2} is in Appendix D. We observe that the only assumptions needed to prove Proposition\,\ref{proposition_2} are (i) the observability of the pair $(C,\mathcal{L})$, and (ii) the generation of $d^2$ linearly independent input vectors $\boldsymbol{\lambda}_{0}^{(\ell)}$, with $\ell=1,\ldots,d^2$. Instead, it does not matter whether the operator $\mathcal{L}$ obeys specific symmetry properties. We can thus conclude that, in case of partial information on the density operator of the network, we need to repeat $d^2$ times the dynamical evolution of the quantum network by starting from $d^2$ independent initial conditions $\boldsymbol{\lambda}_{0}$, each of them corresponding to a specific initial state $\rho_0 \equiv |\psi_0\rangle\!\langle \psi_0| \in \mathbb{C}^{d\times d}$. A quite trivial choice could be to take $\rho_0$ equal to $|k\rangle\!\langle j| \equiv \boldsymbol{e}_{k}\boldsymbol{e}_{j}^{T}$ for $k,j=1,2,\ldots,d$, where $\boldsymbol{e}_{k}$ denotes the $k$-th standard $\mathbb{R}^d$ basis vector. However, let us note that the operators $|k\rangle\!\langle j|$, with $k \neq j$, are not Hermitian and do not have unit trace. Thus, these states are not physical, in the sense that they cannot be experimentally prepared. To overcome this issue, one can decompose $|k\rangle\!\langle j|$ as a function of the fixed set of states $\left\{|k\rangle,\,|j\rangle,\,|+\rangle \equiv (|k\rangle + |j\rangle)/\sqrt{2},\,|+_{y}\rangle \equiv (|k\rangle + i|j\rangle)/\sqrt{2}\right\}$, namely\,\cite{TameNJP2007} 
\begin{equation*}
    |k\rangle\!\langle j| \equiv |+\rangle\!\langle +| + i|+_{y}\rangle\!\langle +_{y}| - \frac{(i+1)}{2}\left(|k\rangle\!\langle k| + |j\rangle\!\langle j|\right),
\end{equation*}
and then exploit the linearity property of any closed quantum dynamics. 

We close this section with some remarks.

\begin{rem}
We have shown that the interaction Hamiltonian (and consequently the topology) of an autonomous quantum dynamical network is identifiable also by properly processing the information coming from partial measurements of $\rho_t$ (in particular, its diagonal elements), provided that $d^2$ linearly independent initial states are prepared. Such initial states have quantum coherence contributions (corresponding to off-diagonal terms in $\rho_0$) between the elements of the basis chosen for their decomposition. Thus, for an accurate reconstruction of the network topology, decreasing the number of intermediate measurements necessarily entails to increase the complexity of the initial quantum state preparation.   
\end{rem}

\begin{rem}
Proposition\,\ref{proposition_2} applies also to quantum networks composed of $N$ interacting $\ell$-levels quantum subsystems. For such systems, indeed, the Liouville–von Neumann equation in linear form can be written as $\dot{\boldsymbol{\lambda}}_t = (\mathcal{L}_{0}+\mathcal{L}_{\rm int})\,\boldsymbol{\lambda}_t$\,, where $\mathcal{L}_0 \equiv -\frac{i}{\hbar}\widetilde{H}_0$ and $\mathcal{L}_{\rm int} \equiv -\frac{i}{\hbar}\widetilde{H}_{\rm int}$. Under the assumptions of Proposition\,\ref{proposition_2} (observability of the pair $(C,\mathcal{L})$ and linearly independent initial states), it is guaranteed that the operator $\mathcal{L}_{0}+\mathcal{L}_{\rm int}$ can be uniquely identified. Therefore, being $\mathcal{L}_0$ fixed and known a-priori, Proposition\,\ref{proposition_2} ensures that also $\mathcal{L}_{\rm int}$ is {\bf uniquely} identifiable.
\end{rem}

\begin{rem}\label{remark_algorithm_2}
Since the quantum network Hamiltonian is the quantity to be determined, a-priori it is unknown if the pair $(C,\mathcal{L})$ is observable. This means that, in general, it is not guaranteed that a possible solution $\hat{M}$ of the topology identification problem is reliable and accurate. Accordingly, for practical purposes, it is worth checking a-posteriori the observability of the pair $(C,\hat{\mathcal{L}})$, with $\hat{\mathcal{L}} \equiv -\frac{i}{\hbar}(\mathbb{I}_{d} \otimes \hat{M} - \hat{M}^T \otimes \mathbb{I}_d)$, which is computed by taking the solution $\hat{M}$ of the identification problem.
\end{rem}

\section{VII. Conclusions}

In this paper, we have addressed the issue of determining under which analytical conditions it is possible to identify the topology of an autonomous quantum network. Specifically, we provide a procedure for the reconstruction of quantum network topologies by using the full-information of the network density operator $\rho_t$, even taken at discrete times. As already discussed in the introductory part of the paper, solutions to this problem could find application in quantum computing and quantum communication, where multiple quantum devices need to be connected in network configurations. In such contexts, information on the network topology is not always available and, as for any other network whose nodes are dynamical systems, even for a quantum network it can happen that one of the nodes looses its functionality or some links suddenly break down. Hence, in all these cases, reliable and easy-to-use tools for topology identification are a prerequisite.  

To achieve our goal in solving the topology reconstruction problem, we have assumed to be able to measure the density operator $\rho_t$ of the network over a given time interval. In particular, we have considered of measuring both all the elements of $\rho_t$ -- for example by means of quantum state tomography -- and only a part of them (just the diagonal elements of $\rho_t$). Though the former case (identification with \emph{full} information) requires more entries of $\rho_t$ to-be-measured than the latter (identification with \emph{partial} information), we provide a sufficient condition for the solvability of the problem by inverting an algebraic commutation relation. This means that, if the identification problem is solvable, the quantum network topology can be reconstructed, in principle, with zero error. The analytical conditions that allow to solve the identification problem with full information have been then converted in a reconstruction algorithm (Algorithm \ref{alg:alg1}). Algorithm \ref{alg:alg1} does not require high computing power and it can be implemented on standard computer facilities. However, one may encounter the issue of computing with few resources (thus, approximately) the operator $P$, obtained by integrating $\rho_t$ within the time interval $[0,\tau]$. The accuracy in calculating this integral depends on the number of time-discrete data points of $\rho_t$ in $[0,\tau]$. In this regard, since at the experimental level the effort in performing lots of measurements is prohibitive, we have numerically verified if the solvability of the topology identification problem with full information is still guaranteed by decreasing the accuracy in computing $P$, namely by using a smaller number of samples. The identification problems remains solvable, though the reconstruction error $\epsilon$ tends to become high, especially if the number of nodes is large or with a too small value of the duration $\tau$ of the quantum network evolution.     

Finally, we have addressed the problem to identify a quantum network topology by using measurements of only the diagonal elements of $\rho_t$. Specifically, here we have shown that the topology of an autonomous quantum network might be reconstructed if the network is observable and is initialized on $d^2$ linearly independent initial states in different runs. 

With others contributions (one can for example refer to the review paper \cite{AltafiniTAC2012}), our paper is one of the attempts to apply analytical results from control theory to quantum mechanical systems. One of our purposes, indeed, is to stimulate contributions that aim at providing exact results in the quantum engineering interdisciplinary field.

Although our analysis has dealt with several facets of the topology identification problem for autonomous quantum networks, it certainly cannot be considered exhaustive. Let us thus mention two possible outlook: \\
(i) Extension of the obtained analytical and numerical results to the case of non-autonomous (or driven) quantum networks. (ii) Solution of the addressed topology identification problem for open quantum networks in interaction with an external environment or other quantum systems. 

\section*{Acknowledgements}

The authors gratefully acknowledge funding from CODYCES, and the University of Florence through the project Q-CODYCES. S.G. and F.C. also acknowledge funding from the Fondazione CR Firenze through the project QUANTUM-AI, and from the European Union’s Horizon 2020 research and innovation programme under Grant Agreement No. 828946 (PATHOS).

\section*{Appendices}
\subsection*{Appendix A -- Interpreting a many-body quantum system as a quantum network}

Let us consider an ensemble of $N$ interacting $\ell$-levels quantum systems, each of them characterized by its own dynamics. Such many-body quantum system is closed, namely it does not interact with the surroundings. As a consequence, the dynamical evolution of the global system is unitary and governed by the total Hamiltonian $H$. As provided by Eq.\,(\ref{eq:H_total}) in the main text, $H$ is given by the sum of the Hamiltonian operators of each subsystem with interaction terms describing the coupling between the subsystems. Formally, $H = \sum_{k}\omega_{k}H_{k} + H_{\rm int}$ with $k$ index on the network subsystems. By assuming (as in the main text) the presence of only two-body interactions and compositions of them at any $t\in[0,\tau]$, the interaction Hamiltonian $H_{\rm int}$ can be further decomposed as the sum of $D \equiv N(N-1)$ coupling terms $A_{k}A_{j}$ with $k \neq j$. Each of them corresponds to a link between two distinct quantum subsystems of the quantum network. Thus, $H_{\rm int} = \sum_{k\neq j}\alpha_{k,j}A_{k}A_{j}$.

Now, as further element, let us observe that also each operator $A_{k}$, with $k=1,\ldots,N$, can be decomposed as the sum of characteristic operators, and specifically as the sum of $\ell^2 -1$ eigen-operators constituting a complete orthonormal basis of the node Hamiltonian operators $H_k$, i.e.,
\begin{equation*}\label{decomposition_A_k}
    A_{k} = \sum_{n=1}^{\ell^2 -1} \beta_{n}A_{k}^{(n)}\,\,\,\, \text{subject to}\,\,\,\,\sum_{n=1}^{\ell^2 -1}A_{k}^{(n)}=\mathbb{I}_d \,.
\end{equation*}
Accordingly, the interaction Hamiltonian $H_{\rm int}$ equals to
\begin{equation*}
    H_{\rm int} = \sum_{k,j=1;\,k \neq j}^{N}
    \sum_{n_1,n_2=1}^{\ell^2 -1}\gamma_{k,j,n_1,n_2}A_{k}^{(n_1)}A_{j}^{(n_2)}
\end{equation*}
where $\gamma_{k,j,n_1,n_2} \equiv \alpha_{k,j}\beta_{n_1}\beta_{n_2}$. For convenience, it is worth introducing the index $m$ that labels the links of the networks. Note that in this analysis both the links $k \rightarrow j$ and $j \rightarrow k$ are distinct elements, generally with a different weight. For example, this means that $m=1$, $m=N-1$ and $m=N(N-1)$ correspond, respectively, to the 2-tuples $(k=1,j=2)$, $(k=1,j=N)$ and $(k=N,j=N-1)$. As a result, $H_{\rm int}$ can be written as the sum of $D_2 \equiv \left(\ell^2 -1\right)^{2}N(N-1)$ elements as in the following relation:
\begin{equation*}\label{eq:H_int_semi-final}
    H_{\rm int} = \sum_{m=1}^{N(N-1)}\sum_{n_1,n_2=1}^{\ell^2 -1}
    \gamma_{m,n_1,n_2}B_{m,n_1,n_2}
\end{equation*}
with $B_{m,n_1,n_2} \equiv A_{k}^{(n_1)}A_{j}^{(n_2)}$, and $n_1,n_2=1,\ldots,\ell^{2}-1$, $m=1,\ldots,N(N-1)$. 

In conclusion, by substituting the expression of $H_{\rm int}$ in $H$, the Liouville-von Neumann equation of a quantum many-body system interpreted as a quantum network is equal to
\begin{equation*}
  \dot{\rho}_t = -\frac{i}{\hbar}\left[(H_0 + \boldsymbol{\gamma}^{T}E\boldsymbol{\gamma})\right] = R_0 -\frac{i}{\hbar}\left[\boldsymbol{\gamma}^{T}E\boldsymbol{\gamma},\rho_t\right]
\end{equation*}
where $R_0$, as in the main text, is completely determined by the knowledge of the local dynamics of each $\ell$-level quantum system, while $\boldsymbol{\gamma}$ and $E$ are provided by the relations
\begin{eqnarray*}
    &&\boldsymbol{\gamma} = \left(\sqrt{\gamma_{1,1,1}}\,\mathbb{I}_{d}\,,\ldots,\sqrt{\gamma_{m,n_1,n_2}}\,\mathbb{I}_{d}\,,\ldots\right)^{T}\in\mathbb{C}^{D_{2}d\times d} \\
    &&E = {\rm diag}\left(\left\{B_{m,n_1,n_2}\right\}\right)\in\mathbb{C}^{D_{2}(d\times d)}\,.
\end{eqnarray*}
The latter formulas for $\boldsymbol{\gamma}$ and $E$ correspond to the ones for $\boldsymbol{\alpha}$ and $C$ in Sec.\,\ref{sec:DQN} of the main text.

\subsection*{Appendix B -- Proof of Theorem \ref{theorem_1}}

Let $M \in \mathcal{A}_\rho$. By integration of the Liouville-von Neumann equation $\dot{\rho}(t) = -\frac{i}{\hbar} [M,\rho(t)]$ between $t=0$ and $t=\tau$, Eq.\,\eqref{skew-Lyapunov} of the main text is obtained. By hypothesis, we have assumed the existence of a unique $\hat{M} \in \mathcal{A}$ satisfying Eq.\,(\ref{skew-Lyapunov}). As a consequence, $|\mathcal{A}_\rho| = 1$, meaning that the set $\mathcal{A}_\rho$ contains only a single element. Thus, since by construction $H \in \mathcal{A}_\rho$, we can conclude that $\mathcal{A}_\rho = \{H\}$.

\subsection*{Appendix C -- Proof of Proposition \ref{proposition_1}}

To demonstrate the ``if'' statement, let assume that $\hat{M}_1, \hat{M}_2 \in \mathcal{A}$ be solutions to Eq.\,\eqref{skew-Lyapunov}. Clearly, we have $(\hat{M}_1-\hat{M}_2) \in \mathcal{A}$. In addition, we obtain
$$
(\hat{M}_1-\hat{M}_2)P-P(\hat{M}_1-\hat{M}_2) = 0
$$
that is, $\hat{M}_1-\hat{M}_2$ commutes with $P$. Thus, by hypothesis we obtain $\hat{M}_1 = \hat{M}_2$, i.e., there is a unique solution to \eqref{skew-Lyapunov} in the set $\mathcal{A}$. Conversely, to prove the ``only if'' statement, let $Z \in \mathcal{A}$ commute with $P$. This implies that $\hat{M} \equiv H + Z \in \mathcal{A}$ is a solution to \eqref{skew-Lyapunov}. Since this solution is unique by hypothesis, we obtain $Z = 0$ that proves the ``only if'' statement and thus the theorem.

\subsection*{Appendix D -- Proof of Proposition \ref{proposition_2}}

Suppose that both the pairs $(C,\mathcal{L})$ and $(C,\overline{\mathcal{L}})$ generate the outputs $\boldsymbol{y}_{\lambda_0^{(\ell)}}(t)$ for $t \in [0,\tau_{\ell}]$, $\tau_{\ell}>0$ and $\ell=1,\ldots,d^2$. Moreover, let us recall that by construction $\dot{\boldsymbol{\lambda}}_t = \mathcal{L}\,\boldsymbol{\lambda}_t$ and $\boldsymbol{y}_t = C\boldsymbol{\lambda}_t$ for any value of $t$. Thus, one can find that
\begin{equation*}
    \dot{\boldsymbol{y}}_{\lambda_0^{(\ell)}}(0) = C\mathcal{L}\boldsymbol{\lambda}_0^{(\ell)} = C\overline{\mathcal{L}}\boldsymbol{\lambda}_0^{(\ell)}.
\end{equation*}
In similar fashion, by computing higher-order derivatives of $\boldsymbol{y}_t$ till the $d^2$-th one, we also get
\begin{equation*}
    C\mathcal{L}^{k}\boldsymbol{\lambda}_0^{(\ell)} = C\overline{\mathcal{L}}^{k}\boldsymbol{\lambda}_0^{(\ell)}
\end{equation*}
for all $k=1,2,\ldots,d^2$. Since the initial states $\boldsymbol{\lambda}_0^{(\ell)}$, with $\ell=1,2,\ldots,d^2$, are linearly independent vectors, it holds that $C\mathcal{L}^{k}=C\overline{\mathcal{L}}^{k}$ for all $k=1,2,\ldots,d^2$. Hence,
\begin{equation*}
    \begin{bmatrix} C \\ C\mathcal{L} \\ \vdots \\ C\mathcal{L}^{d^2 - 1} \end{bmatrix}\mathcal{L} = \begin{bmatrix} C \\ C\overline{\mathcal{L}} \\ \vdots \\ C\overline{\mathcal{L}}^{d^2 - 1} \end{bmatrix}\overline{\mathcal{L}} = \begin{bmatrix} C \\ C\mathcal{L} \\ \vdots \\ C\mathcal{L}^{d^2 - 1} \end{bmatrix}\overline{\mathcal{L}}.
\end{equation*}
Equivalently, 
\begin{equation*}
    \begin{bmatrix} C \\ C\mathcal{L} \\ \vdots \\ C\mathcal{L}^{d^2 - 1} \end{bmatrix}(\mathcal{L} - \overline{\mathcal{L}}) = 0.
\end{equation*}
Finally, by resorting to the assumption that the pair $(C,\mathcal{L})$ is observable, we can conclude that $\mathcal{L} = \overline{\mathcal{L}}$, thus proving the Proposition. 

%% ---------------------------------------
%% Bibliography
%% ---------------------------------------

\end{document}